# Sthread: In-Vivo Model Checking of Multithreaded Programs

## Gene Cooperman[a] and Martin Quinson[b]

a   Khoury College of Computer Sciences, Northeastern University, United States, Boston, USA
b   Univ. Rennes, Inria, CNRS, IRISA, Rennes, France

**Abstract**   This work strives to make formal verification of POSIX multithreaded programs easily accessible to general programmers. Sthread operates directly on multithreaded C/C++ programs, without the need for an intermediate formal model. Sthread is in-vivo in that it provides a drop-in replacement for the pthread library, and operates directly on the compiled target executable and application libraries. There is no compiler-generated intermediate representation. The system calls in the application remain unaltered. Optionally, the programmer can add a small amount of additional native C code to include assertions based on the user's algorithm, declarations of shared memory regions, and progress/liveness conditions. The work has two important motivations: (i) It can be used to verify correctness of a concurrent algorithm being implemented with multithreading; and (ii) it can also be used pedagogically to provide immediate feedback to students learning either to employ POSIX threads system calls or to implement multithreaded algorithms.

This work represents the first example of in-vivo model checking operating directly on the standard multithreaded executable and its libraries, without the aid of a compiler-generated intermediate representation. Sthread leverages the open-source SimGrid libraries, and will eventually be integrated into SimGrid. Sthread employs a non-preemptive model in which thread context switches occur only at multithreaded system calls (e.g., mutex, semaphore) or before accesses to shared memory regions. The emphasis is on finding "algorithmic bugs" (bugs in an original algorithm, implemented as POSIX threads and shared memory regions. This work is in contrast to Context-Bounded Analysis (CBA), which assumes a preemptive model for threads, and emphasizes implementation bugs such as buffer overruns and write-after-free for memory allocation. In particular, the Sthread in-vivo approach has strong future potential for pedagogy, by providing immediate feedback to students who are first learning the correct use of Pthreads system calls in implementation of concurrent algorithms based on multithreading.

# The Art, Science, and Engineering of Programming

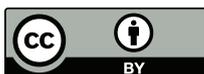

**Sthread: In-Vivo Model Checking of Multithreaded Programs**

## 1 Introduction

Sthread is an in-vivo, explicit-state model checker used to find race conditions, deadlock, assertion failures, and other bugs associated with multithreaded programs. Sthread executes directly on C/C++ code that implements concurrent algorithms using the POSIX multithreaded system calls (mutex, semaphore, etc.) and shared memory regions. The user's C/C++ code is compiled as usual by the user's preferred compiler, but Sthread replaces the Linux pthread library with an Sthread/SimGrid library and the user's C/C++ code is re-compiled with an sthread include file instead of the traditional pthread include file. As with all formal verification approaches, Sthread can capture rare bugs such as data races, which can be missed by test suites or unit testing.

Sthread is based on recent advances in the SimGrid software, along with a new SimGrid interface, sthread. SimGrid is an explicit-state model checker that provides a SimGrid library that is linked to the target code. Routines in the target code are redirected to a library that makes "simcalls" to the SimGrid thread scheduler. Typically, a simcall represents the end of a transition between model states, and the process memory itself is sued to create the formal model state of the SimGrid model checker. Transitions between states in the formal model are defined by choosing a single thread and running it until either a thread-based system call (e.g., mutex, semaphore) is made or else until an access to a shared memory region occurs. It is implicitly assumed that the code executed in transitions is deterministic and non-blocking. Hence, each transition from a given state corresponds to a unique thread that is ready to run.

The contributions of this work are:

1. This is the first example of in-vivo model checking operating directly on a standard multithreaded executable (where the process memory is the state).
2. By limiting thread context switches to POSIX system calls, Sthread is better able to limit the exponential explosion of states and explore more deeply into the execution of a program. This is suitable for checking the algorithm used by an implementation, both in production code and for checking student-generated code for pedagogical purposes.
3. Sthread allows the programmer to add simple C/C++ code (e.g., sched_yield) to annotate the algorithmic use of shared memory, or the addition of C/C++ variables to verify application-specific liveness properties. The power of this approach is in keeping the Sthread implementation small and efficient, in contrast to ambitious approaches employing more costly models of thread preemption and general application-independent liveness properties.
4. Broad coverage of target codes is demonstrated through a series of examples to check a broad variety of multithreaded paradigms: mutexes, semaphores, shared memory regions, lock-free algorithms, and other cases.

A consequence of the above contributions is that Sthread's approach is both language-independent and compiler-independent. Sthread supports native executables and native libraries. Hence, Sthread can easily support other programming languages, and even other communication libraries. In the case of distributed communication, this was previously shown through SMPI/SimGrid's support for MPI [11].

The Sthread/SimGrid approach (in which process memory is the state) is related to Context-Bounded Analysis (CBA). However, CBA is typically employed to model preemptive context switching. A small bound is placed on the total number of thread context switches, and a thread context switch may occur between any two instructions in the target program. In contrast, Sthread allows a context switch only at a POSIX thread system call, or before an access to a shared memory region.

The organization of the paper follows. Section 2 presents an overview of the work. Section 4 presents the design and implementation. Section 3 illustrates the use of Sthread with a concrete example wherein a race condition is discovered and a deterministic thread schedule is produced to enable easy debugging. Section 5 presents case studies demonstrating the use of Sthread. Finally, section 6 describes the related work, and section 7 presents the conclusion and describes future work.

## 2 Overview

This work addresses bugs in multithreaded implementations of concurrent algorithms. The life cycle of such a bug passes through three phases:

1. identification that a bug exists
2. diagnosis through an explicit execution trace
3. creation of a bug fix — ideally, through the use of deterministic replay in a debugging environment

**Phase 1: Identification of a bug**  Classical identification of a bug relies on a program crash, a hang (deadlock or livelock), an assert failure, or the return of an incorrect result (which can be caught by an assertion). The use of formal verification ensures that a schedule will be found that exposes such a failure mode (subject to the exponential explosion of states). Sthread uses explicit-state model checking to find such a schedule (subject to the model of non-preemption of threads).

**Phase 2: Diagnosis**  Classically, the second phase is addressed by deterministic record-replay [10, 22, 26, 28]. Reversible debuggers [14, 31] are another option, but they are less suitable for multithreaded programs since they do not provide guarantees of deterministic replay. For record-replay, if an execution enters a failure mode while recording, then a deterministic replay of that execution is provided. In the case of Sthread, the identification of the bug in Phase 1 is found through a deterministic thread execution schedule. Hence, a deterministic replay is immediately available.

**Phase 3: Bug fix**  Given the existence of a deterministic replay schedule, it is fairly simple to replay the execution trace within the environment of a debugger. Sthread, like other approaches, provides this opportunity.

■ **Listing 1** Code snippet for mutex-deadlock.cpp

```cpp
#include <sthread.h>

pthread_mutex_t mutex1 = PTHREAD_MUTEX_INITIALIZER;
pthread_mutex_t mutex2 = PTHREAD_MUTEX_INITIALIZER;

static void* thread2_start() {
  while (true) {
    pthread_mutex_lock(&mutex2);
    pthread_mutex_lock(&mutex1);
    pthread_mutex_unlock(&mutex1);
    pthread_mutex_unlock(&mutex2);
  }
  return NULL;
}

int main(int argc, char* argv[])
{
  pthread_t thread2;
  pthread_create(&thread2, NULL, thread2_start, NULL);

  while (true) {
    pthread_mutex_lock(&mutex1);
    pthread_mutex_lock(&mutex2);
    pthread_mutex_unlock(&mutex2);
    pthread_mutex_unlock(&mutex1);
  }

  return 0;
}
```

## 3 A First Example: Deadlock with Mutexes

In order to make the ideas more concrete, it is best to start with a simple example showing the use of Sthread. Consider the following classic example of deadlock based on mutexes.

The code in table 1 is compiled using sthreadc++, a simple wrapper script for the native C++ compiler that employs the SimGrid library as a drop-in replacement for the Linux pthread library. After compiling, the resulting binary, mc-mutex-deadlock, is executed using simgrid-mc, the SimGrid Model Checker.

```
bin/simgrid-mc ./mc-mutex-deadlock examples/platforms/small_platform.xml \
  "--log=root.fmt:[%10.6r]%e(%i:%P@%h)%e%m%n" \
 --cfg=model-check/reduction:none --cfg=model-check/max-depth:10
```

Note particularly the setting of max-depth to 10. This means that Sthread will explore the execution along all possible thread schedules, and it will terminate the exploration along a given thread schedule when more than 10 calls to pthread_mutex_lock or pthread_mutex_unlock have occurred. Sthread then reports the following thread schedule as producing deadlock.

```
[ 0.000000] (0:maestro@) /!\ Max depth reached ! /!\
[ 0.000000] (0:maestro@) **************************
[ 0.000000] (0:maestro@) *** DEADLOCK DETECTED ***
[ 0.000000] (0:maestro@) **************************
[ 0.000000] (0:maestro@) Counter-example execution trace:
[ 0.000000] (0:maestro@) Path = 1;1;1;1;1;1;1;1;2
```

The interpretation of this "Path" is simple. Assume that Thread 1 executes by locking and unlocking mutexes during 9 transitions. Since each iteration of the "while" loop invokes 4 such calls, the 9$^{th}$ call results in thread 1 acquiring mutex1. At this point, thread 2 executes to acquire mutex2. It is then impossible for either thread to continue and acquire the remaining mutex.

Hence, the path given by Sthread provides a deterministic thread schedule leading to deadlock. Sthread has not only identified that the code contains a bug due to a race condition, but Sthread has produced a deterministic thread schedule suitable for deterministic replay. Deterministic replay offers the original programmer the opportunity not only for bug diagnosis (Phase 2 from the Overview), but also for testing candidates for bug fixes (Phase 3 from the Overview).

## 4 Design and Implementation

We break up the discussion into two parts: the overall design of Sthread/SimGrid and some details of its implementation.

**Design of Sthread:** Sthread is based on the SimGrid model checker. SimGrid was originally designed for checking distributed algorithms and protocols. It was later adapted for model checking of MPI programs. SimGrid operates by providing a communication library that replaces the native communication library. in the case of distributed programming, SimGrid provides a replacement for the native inter-process communication (IPC) library. In the case of MPI, SimGrid provides its own library (as part of its "smpi" component) to replace the native MPI library. Further information and references for SimGrid are contained in section 6.

Sthread has extended the SimGrid built-in utilities for mutexes, semaphores and other multi-threaded constructs in order to create a Pthread (POSIX threads) component and interface that is compatible with the standard Linux pthread library. Thus, instead of linking against the Linux pthread library, Sthread links against the SimGrid library, with a thin compatibility interface provided by the Sthread component.

While it is beyond the scope of this work to fully review model checking, its essence can be easily deduced from figure 1. That figure illustrates the internal states that are explored by Sthread/SimGrid as it executes the binary based on table 1 of the previous section.

Each transition can be labeled by a unique thread that executes during a transition between states. Of course, several threads may execute concurrently in a multi-core CPU or due to context switching. However, at least when there are no interactions through shared memory, which is the case for table 1, then the idealized execution presented here is semantically equivalent to the actual execution.

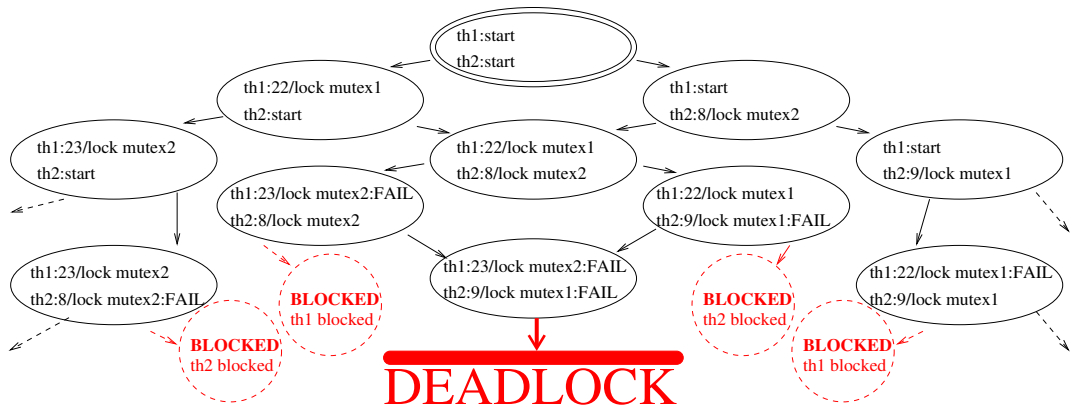

**Figure 1** A model checker explores all reachable, distinct *states* of a program or formal model. Each state has zero or more outgoing *transitions* to other states. Leftward arrows indicate that thread 1 was scheduled next, and rightward arrows indicate that thread 2 was scheduled next. If a mutex lock is temporarily blocked, this is indicated by "FAIL" in the figure. Otherwise, all lock attempts are assumed to succeed. The figure is based on table 1.

A path in figure 1 corresponds to a *thread schedule*. Since each transition of the path can be labeled by a unique thread, Sthread produces a sequence of thread ids corresponding to that path. It is this thread schedule that enables deterministic replay to enable easy debugging, as referenced in the "Diagnosis" phase in section 2.

In addition to deadlock (no outgoing transitions from a state) that is illustrated in figure 1, Sthread can also detect rare program crashes (transition sequences that lead to a crash in the program), program assertion failures (transitions whose execution causes a program assertion to fail), livelock (given by a subset of the states such that there is no transition leaving the subset), and errors in the concurrent use of shared memory by multiple threads.

In order for Sthread to detect livelock and shared memory concurrency errors, the programmer must introduce additional code. In the case of livelock, additional code to annotate a progress condition must be added. For an example of handling livelock, see section 5.3. For an example of debugging shared memory concurrency, see section 5.1.

Of course, Sthread, like all model checkers, suffers from the problem of an exponential explosion of states. It is not claimed that Sthread can discover all race conditions in some category of POSIX multi-threaded programs.

Our goal here is simply to show that Sthread discovers race conditions in many interesting multi-threaded programs, including some of interest in real-world programs. In practice, it may be necessary to isolate the code of interest, and to simplify it in order to escape from an exponential explosion of states. Many approaches for alleviating that explosion of states are well-known in the literature, but a discussion of such strategies is beyond the scope of this work.

**Key SimGrid feature: SimGrid model states are based on a snapshot of process memory.**
Before continuing on to the implementation of Sthread, there is one additional observation to be made about SimGrid. The model state employed by SimGrid is literally the same as the memory of the Linux process. SimGrid has a variety of heuristics to ignore "garbage memory" (e.g., uninitialized memory returned by "malloc", or padding within a "struct"). A new model state is recognized as equivalent to a previous model state if the significant process memory is the same. This recognition of previously seen model states helps to control the exponential explosion of states. SimGrid also employs additional techniques from formal verification to control the exponential explosion.

**Implementation of Sthread:** Internally, Sthread currently uses macros in a file sthread.h for simple source-to-source transformations in order to introduce simcalls to the SimGrid scheduler. Two scripts, sthreadcc and sthreadc++, serve as wrappers by which to redirect the line #include <pthread.h> to an sthread-based directory that in turn will include the sthread.h file. Future versions will directly preprocess the programmer's target source code, and also interpose on calls to POSIX threads system calls at runtime. The latter ability will allow Sthread to also support application libraries that make use of POSIX threads.

The sthread.h include file translates portions of the programmer's source code into a form that is compatible with the underlying SimGrid. As an example, main is expanded into a new code for main, which then initializes the underlying SimGrid engine and finally delegates control to an auxiliary function, primary_thread, which executes the programmer's application code. Similarly, the standard system calls for mutexes are expanded into calls appropriate for SimGrid. Listing 2 shows excerpts of sthread.h that manage this transformation.

In the expanded programmer's code, a call such as pthread_mutex_lock(mymutex_ptr) is expanded into (*mymutex_ptr)->lock. The latter expression is native to SimGrid, and creates a "simcall" down to SimGrid's scheduler so that SimGrid can explore all outgoing transitions (all context switches to a thread that is ready to run).

More generally, the POSIX system calls each generate a simcall that marks the end of a state transition. The new state is defined by the state of process memory, in keeping with the SimGrid implementation [20]. At that time, SimGrid's schedule can schedule any ready-to-run thread. In keeping with explicit-state model checkers, all paths (corresponding all ready-to-run threads) are explored.

In addition to the standard POSIX system calls, the programmer may insert into the code a call to sched_yield(). This also gets translated into a SimGrid primitive that allows SimGrid to schedule other threads. The programmer is encouraged to place sched_yield() in front of any access to shared memory by the application, so that the current thread may pause and later "see" any modifications to the shared memory by other threads.

Note that a transition in Sthread is assumed to consist of a sequence of instructions that are deterministic and non-blocking. This is in keeping with the philosophy that the execution of Sthread should reflect the underlying concurrent algorithm that is being implemented by the multithreaded program.

**Sthread: In-Vivo Model Checking of Multithreaded Programs**

■ **Listing 2** Code snippet for sthread.h

```
1  #define main(...) \
2  main(__VA_ARGS__) { \
3    void main_simgrid(__VA_ARGS__); \
4    main_simgrid(argc, argv); \
5  } \
6  \
7  int primary_thread(__VA_ARGS__)
8
9  // primary_thread() is executed by
       ↪ main_simgrid()
10 int primary_thread(int argc, char* argv[]);
11
12 std::vector<simgrid::s4u::Host*> hosts;
13
14 inline void main_simgrid(int argc, char* argv
       ↪ [])
15 {
16   simgrid::s4u::Engine e(&argc, argv);
17   xbt_assert(argc > 1, "Usage: %s
       ↪ platform_file\n", argv[0]);
18
19   e.load_platform(argv[1]);
20   hosts = e.get_all_hosts();
21   xbt_assert(hosts.size() >= 1, "This example
       ↪ requires at least 1 hosts");
22
23   simgrid::s4u::Actor::create("primary thread",
       ↪ hosts[0], &primary_thread, argc,
       ↪ argv);
24
25   e.run();
26 }
27
28 //====================
29 // Code snippets to translate mutex to
30 // the underlying SimGrid constructs
31 #undef pthread_mutex_t
32 #define pthread_mutex_t simgrid::s4u::
       ↪ MutexPtr
33 #undef PTHREAD_MUTEX_INITIALIZER
34 #define PTHREAD_MUTEX_INITIALIZER simgrid
       ↪ ::s4u::Mutex::create()
35
36 // Ensure that: pthread_mutex_lock(&
       ↪ mymutex) translates correctly for
       ↪ SimGrid:
37 #undef pthread_mutex_lock
38 #define pthread_mutex_lock(mymutex_ptr) (*
       ↪ mymutex_ptr)->lock()
39 #undef pthread_mutex_unlock
40 #define pthread_mutex_unlock(mymutex_ptr
       ↪ ) (*mymutex_ptr)->unlock()
```

Once the POSIX system calls have been translated for SimGrid, Sthread then relies on SimGrid to produce execution traces. This is best illustrated in reviewing the case studies (section 5).

Next, in order to support a debugging environment for replaying Sthread traces (see Phase 3 in section 2), a gdbinit file for GDB is generated from the execution trace for some failure mode. The gdbinit file is then "sourced" into a GDB session and used to deterministically reproduce the execution trace. In order to manage the replay of a specified thread at each step, GDB's scheduler-locking mode is used. (A line set scheduler-locking on is inserted into the gdbinit file.) Further a GDB variable, $counter is initialized to 1 and is incremented after each "continue" call to GDB by the programmer. Incrementing a GDB variable is as easy as set $counter=$counter+1. A line is inserted into the GDB file of the form, for example:

(gdb) set $trace={2,1,1,2}

where {2,1,1,2} might be the execution trace captured by Sthread. Finally, GDB-based Python commands are used to capture control after each call to "continue" and to then execute:

(gdb) thread $trace[$counter]

**Listing 3** Code snippet for hello-shared-memory.cpp

```
1  #include <sthread.h>
2
3  bool val1=false;
4  bool val2=false;
5
6  // A POSIX thread takes an argument of type '
       ↪ void*'
7  static void* thread2_start() {
8    // yield before reading shared variable '
         ↪ val1'
9    sched_yield();
10   if (not val1) {
11     // Yield before writing to shared variable
           ↪ 'val2'
12     sched_yield();
13     val2 = true;
14   }
15   assert(not (val1 and val2));
16
17   return NULL;
18 }

19
20 int main(int argc, char* argv[])
21 {
22   pthread_t thread2;
23   pthread_create(&thread2, NULL,
         ↪ thread2_start, NULL);
24
25   // yield before reading shared variable '
         ↪ val2'
26   sched_yield();
27   if (not val2) {
28     // yield before writing to shared variable
           ↪ 'val1'
29     sched_yield();
30     val1 = true;
31   }
32   assert(not (val1 and val2));
33
34   return 0;
35 }
```

## 5 Pedagogical Examples

This next section includes the following examples, selected from a hypothetical pedagogical tutorial in order of difficulty.

1. Two threads refer to each other (example for sequential consistency)
2. Deadlock using Mutexes
3. Priority inversion (Mars Pathfinder)
4. ABA problem for lock-free algorithms (from DMTCP code)

### 5.1 Sthread Hello world: Handling Shared Memory

We begin by exhibiting one of the simplest multithreaded applications. Pedagogically, it is important in showing a flaw in a naive leader election implementation. Naive intuition assumes that the "if" statement is atomic. Either the opposite thread's variable remains "false", or we detect that it is "true" and we leave our own variable at "false". The assert statements in table 3 (lines 15 and 32) show that this naive intuition is wrong. It is possible for *both* threads to win the leader election, even though each thread sets its own variable to "true" while assuming that the opposite thread's variable is "false".

Unlike the mutex example presented in section 5.2, Sthread requires the programmer to explicitly add the POSIX system call, sched_yield. Note that this addition does not change the semantics of the original program.

The sched_yield call in the listing is used to mark the beginning of each access to shared memory. This results in a simcall to the SimGrid library so that SimGrid may simulate a thread context switch at this point in the code. Without this Sthread/SimGrid would assume that there are no context switches at all. There is no need to also place sched_yield in front of Statements that do not access shared memory, since an *additional* thread context switch at that point in the code cannot change the semantics of the program.

In principle, such a sched_yield call must be placed in front of every primitive operation that can access memory shared by more than one thread. But in practice, it often suffices to place the call in front of an entire read or write of a block of memory.

Arguably, it places a significant burden on the programmer to identify all program variables that may be accessed by more than one thread. But in practice, a programmer should understand the algorithm used in the code, which implies a knowledge of which variables may be shared by multiple threads. Of course, there are exceptional cases such as buffer overruns, where the programmer is not aware of memory sharing between threads. Sthread is not able to find such exceptional bugs as buffer overruns or writes to arbitrary addresses in memory.

Finally, we run Sthread on the code, yielding the following output.

```
[ 0.000000] (0:maestro@) **************************
[ 0.000000] (0:maestro@) *** PROPERTY NOT VALID ***
[ 0.000000] (0:maestro@) **************************
[ 0.000000] (0:maestro@) Counter-example execution trace:
[ 0.000000] (0:maestro@) Path = 1;2;1
[ 0.000000] (0:maestro@) Expanded states = 5
[ 0.000000] (0:maestro@) Visited states = 7
[ 0.000000] (0:maestro@) Executed transitions = 6
```

In this case, the path is 1;2;1. Thread 1 completes a call to sched_yield and stops at line 29. Thread 2 completes a call to sched_yield and stops at line 12. Thread 1 completes a second call to sched_yield, sets val2 to true, and validates the assertion not (val1 and val2) since val2 is still false. After this, thread 1 has finished, and thread 2 would then execute a second sched_yield, set val2 to true and try to validate the assertion not (val1 and val2). But that assertion now fails. So, no more transitions by thread 1 are possible, and the state transition by thread 2 produces "PROPERTY NOT VALID" on assertion failure.

### 5.2 Deadlock with Mutexes: Deadly Embrace

The example of deadlock with mutexes would often come next in a pedagogical sequence. This example was already seen in section 3, where table 1 was presented.

We next explore Sthread further by varying the max-depth. We first set max-depth to 1 (at most one scheduling point).

```
bin/simgrid-mc ./mc-mutex-deadlock examples/platforms/small_platform.xml \
  "--log=root.fmt:[%10.6r]%e(%i:%P@%h)%e%m%n" \
 --cfg=model-check/reduction:none --cfg=model-check/max-depth:1
```

Sthread correctly reports that if we stop after the first mutex_lock, then no property violation and no deadlock is found:

```
[ 0.000000] (0:maestro@) /!\ Max depth reached ! /!\
[ 0.000000] (0:maestro@) No property violation found.
[ 0.000000] (0:maestro@) Expanded states = 3
[ 0.000000] (0:maestro@) Visited states = 4
[ 0.000000] (0:maestro@) Executed transitions = 2
```

As soon as we raise max-depth to 2, we see the minimal example of deadlock:

```
[ 0.000000] (0:maestro@) /!\ Max depth reached ! /!\
[ 0.000000] (0:maestro@) **************************
[ 0.000000] (0:maestro@) *** DEADLOCK DETECTED ***
[ 0.000000] (0:maestro@) **************************
[ 0.000000] (0:maestro@) Counter-example execution trace:
[ 0.000000] (0:maestro@) Path = 1;2
[ 0.000000] (0:maestro@) Expanded states = 4
[ 0.000000] (0:maestro@) Visited states = 6
[ 0.000000] (0:maestro@) Executed transitions = 4
```

Raising max-depth to 10, results in

```
[ 0.000000] (0:maestro@) Path = 1;1;1;1;1;1;1;1;2
```

which was discussed in section 3.

In fact, the Python script sthread-check calls on the simgrid-mc executable multiple times in a binary search to discover the smallest max-depth that results in a deadlock or other violation. The default depth (configurable) is currently set to 1000.

### 5.3 Priority Inversion and Liveness Properties

Next, an example of priority inversion is shown. This example of priority inversion with three threads reflects a famous historical case involving the Mars Pathfinder. That bug was discovered and patched by engineers back on Earth.

In a case of priority inversion, a low-priority thread acquires a mutex lock (a bus_lock in the following code), but releases that mutex lock to a high-priority thread on demand. The low-priority thread then blocks on the mutex. Unfortunately, a medium-priority process shares a second resource (protected by a task_lock in the example code below) with the low-priority thread. In this case, the medium-priority lock acquires the task_lock, forcing the low-priority thread to block. The high-priority thread requests that the low-priority thread release the bus_lock, but the low-priority thread is not responding. The result is livelock since only the low-priority thread is blocked. In the code below, thread 1 is the high-priority thread, thread 2 is the medium-priority thread, and thread 3 is the low-priority thread.

In order for Sthread to analyze this bug, the programmer must add a progress condition into the code. SimGrid offers a simple utility, Promela, for defining liveness conditions. This is typically used for packages like SimGrid that employ Büchi automata internally. The details are described in the SimGrid documentation and elsewhere, but are outside the scope of this work.

# Sthread: In-Vivo Model Checking of Multithreaded Programs

Promela is used as a front end in order to pass a certain state machine for use in the internals of SimGrid. From the viewpoint of the Sthread user, the code related to Promela is:

1. Lines 106 and line 107: MC_automaton_new_propositional_symbol_pointer(): These declare cs and r as parameters associated with the liveness condition.
2. Lines 20, 21, 28, 29, 34, 41, 42: occurrences of cs and r.

The variable r is intended to denote that a "request" is active. The variable cs is intended to denote that a thread is in the "critical section". These variables are associated with thread 1, the high priority thread. The Promela-based liveness condition is $G(r-> Fcs)$, with the intended meaning in LTL formal logic: "It is always true that if a request is made then eventually the critical section will be entered." They define a progress condition. Thread 1 is setting a variable, priority1_interrupt to true before attempting to acquire the two mutex locks. The liveness condition fails if there is a thread schedule such that eventually neither cs nor r change their state. For details, one must consult the SimGrid documentation.

■ **Listing 4** Code snippet for priority inversion. A low priority thread acquires a bus lock. It is interrupted by a medium priority thread, but the low priority thread continues to hold the bus lock. The high priority thread then tries to gain control, but since the low priority thread is waiting on the medium thread, it cannot gain control and give up the bus lock to the high priority thread.

```
/*** Non-deterministic thread ordering ***
/* Priority inversion: Thread1 (primary_thread)
/* has highest priority, followed by thread2,
/* and then thread3. A famous example of this
/* occurred for the Mars Pathfinder:
/* https://www.rapitasystems.com/blog/what-really-
    ↪ happened-to-the-software-on-the-mars-
    ↪ pathfinder-spacecraft */
/*****************************************

#include <sthread.h>

pthread_mutex_t task_lock =
    ↪ PTHREAD_MUTEX_INITIALIZER;
pthread_mutex_t bus_lock =
    ↪ PTHREAD_MUTEX_INITIALIZER;

bool priority1_interrupt = false;
bool priority2_interrupt = false;
bool task1_event = false;
bool task2_event = false;
bool task3_event = false;

int r=0;
int cs=0;

// Thread1: high priority
static void* thread1_start() {
  while (true) {
    sched_yield(); // prior to reading a shared variable
    if (task3_event) {
      r=1;
      cs=0;
      sched_yield(); // prior to writing a shared variable
      priority1_interrupt = true; // Ask thread3, release
          ↪ task_lock, bus_lock.
      pthread_mutex_lock(&task_lock);
      pthread_mutex_lock(&bus_lock);
      cs=1;
      // do_task1();
      // If we could create an extra state here, we would
          ↪ reset cs to '0' here.
      sched_yield(); // prior to writing a shared variable
      priority1_interrupt = false; // thread2 or thread3
          ↪ can resume now
      pthread_mutex_unlock(&bus_lock);
      pthread_mutex_unlock(&task_lock);
      cs=0;
      r=0;
    }
  }

  return NULL;
}

// Thread2: medium priority
static void* thread2_start() {
  while (true) {
    sched_yield(); // prior to reading a shared variable
    while (task2_event) {
      sched_yield(); // prior to writing a shared variable
      priority2_interrupt = true; // Ask thread3 to release
          ↪ task_lock.
      pthread_mutex_lock(&task_lock);
      // do_task2();
      // In principle, this priority1_interrupt could arrive
          ↪ in the middle
      // of do_task2(), above.
      sched_yield(); // prior to reading a shared variable
      if (priority1_interrupt) {
        pthread_mutex_unlock(&task_lock);
        // Perhaps sleep for a while, so that thread1 can
            ↪ do its task
```

```
      pthread_mutex_lock(&task_lock);
    }
    // do_task2(): do rest of task2 after interrupt
    sched_yield(); // prior to writing a shared variable
    task2_event = false; // Done executing. Check if
        ↪ there is more to do.
    priority2_interrupt = false; // thread3 can resume
        ↪ now
    sched_yield(); // prior to reading a shared variable
    pthread_mutex_unlock(&task_lock);
  }
}

  return NULL;
}

// Thread3: lowest priority
static void* thread3_start() {
  sched_yield(); // prior to reading a shared variable
  while (task3_event) {
    pthread_mutex_lock(&task_lock);
    pthread_mutex_lock(&bus_lock); // This was a
        ↪ semaphore in original example.
    sched_yield(); // prior to reading a shared variable
    if (priority1_interrupt or priority2_interrupt) {
      // Unlock; a higher priority task wants the lock
      pthread_mutex_unlock(&task_lock);
      while (1) {
        sched_yield(); // prior to reading a shared
            ↪ variable
        if (not priority1_interrupt and not
            ↪ priority2_interrupt) { break; }
      }
      pthread_mutex_lock(&task_lock);
    }
    // do_task2();
    sched_yield(); // prior to writing a shared variable
```

```
    task3_event = false; // finished task
    pthread_mutex_unlock(&bus_lock);
    pthread_mutex_unlock(&task_lock);
  }

  return NULL;
}

int main(int argc, char* argv[])
{
  MC_automaton_new_propositional_symbol_pointer("r",
      ↪ &r);
  MC_automaton_new_propositional_symbol_pointer("cs
      ↪ ", &cs);
  // MC_ignore(&(status.count), sizeof(status.count));

  pthread_t thread1, thread2, thread3;
  pthread_create(&thread1, NULL, thread1_start, NULL);
  pthread_create(&thread2, NULL, thread2_start, NULL);
  pthread_create(&thread2, NULL, thread3_start, NULL);

  for (int i=0; i < 10; i++) {
    // In principle, we should be generating these events
    // at random, but this is close enough, and will still
    // generate the same bug.
    sched_yield();
    task1_event = true;
    sched_yield();
    task2_event = true;
    sched_yield();
    task3_event = true;
    sched_yield();
  }

  return 0;
}
```

This code uses the same liveness property as in section 5.2, but this time, the file is called promela-priority-inversion. As expected, this code fails, and in this case with the execution trace 1;1;…;1;2;2;0;2;2;2;.

### 5.4 The ABA Problem: A Long-Standing Challenge

The next pedagogical example concerns an extremely subtle bug in lock-free implementations, known as the ABA problem. We pause here to to provide some background for the ABA problem. In subsection 5.5, we then continue by presenting a flawed lock-free implementation as it appeared in a project of the first author. We show how Sthread automatically discovers a subtle race condition, and presents a thread schedule for deterministic analysis of the bug.

Figure 2 describes the well-known ABA problem. The phrase "pop(A)" is used to indicate a call to "pop()" that returns the item A.

In figure 2, assume that we wish to allocate and deallocate items to or from the head of a stack implemented as a linked list. In step (a) of the diagram, the current thread wishes to pop(A), and set the pointer x to point to item B. However, the thread must ensure that no outside thread will come in and itself pop item A from the stack.

**Sthread: In-Vivo Model Checking of Multithreaded Programs**

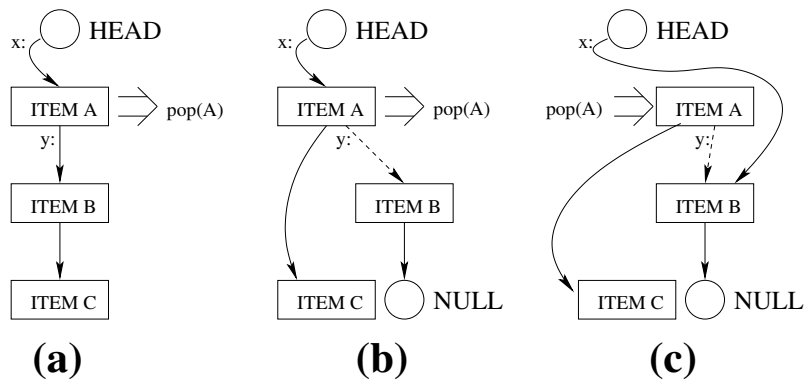

**Figure 2** ABA Problem [33]: a thread intends to pop item A in step (a); a concurrent, second thread "steals" item B from the stack in step (b); this causes the original thread to do the wrong thing in step (c).

In order to implement the intended operations, it is necessary to make recourse to an atomic operation at the assembly level such as test-and-set. As will be seen later in listing 5, the GCC toolchain and other compilers supports a similar atomic built-in statement:

```
bool __sync_bool_compare_and_swap (type *ptr, type expected, type desired)
```

where "type" is any common primitive type (e.g., int) supported by the CPU.

This is encoded as:

```
__sync_bool_compare_and_swap (&x, x, y);
```

The atomic built-in has the effect of the following snippet of code, except that it is atomic, and no other thread can interfere between the test for expected, and setting x to desired.

```
if (*ptr == expected) { *ptr = desired; }
```

So, we can implement "deallocate" as:

```
~~ while (SUCCESS != __sync_bool_compare_and_swap (&x, x, y)) { };
```

Assume further that in step (a), the main thread has loaded the pointers a and b into registers.

Unfortunately, in step (b), a second thread then arrives and executes:

```
tmp = pop(A);
pop(B);
push(tmp); // Place A at head of stack again.
```

Next, in step (c), the original thread finally executes:

```
__sync_bool_compare_and_swap (&x, x, y);
```

But since x and y were stored in registers, their values were not modified by step (b). So, we are left with the diagram in step (c) in which x has been set to point to *y. But this does not have the desired effect. Item C and the rest of the stack have now been removed from the stack, and are no longer reachable from the head of the stack.

This example happened to the first author, in his work with the DMTCP project [2]. DMTCP (Distributed MultiThreaded CheckPointing) is a widely used package for transparent checkpoint-restart of Linux applications. The DMTCP developers had written a lock-free memory allocator for the sake of efficiency. The developers, including this first author, had not observed any bugs associated with this code during the 2-1/2 years after the DMTCP code had been written. During this time, it is estimated that thousands of users of DMTCP were running this lock-free code without triggering a bug. Then a new, reproducible bug manifested itself in an unusual target application that was being checkpointed by DMTCP. The application created and destroyed thousands of threads per second, while allocating and deallocating memory. The original lock-free implementation in DMTCP is provided verbatim in listing 5 as part of the discussion of section 5.5.

It took the developers of DMTCP a week to identify this bug, known as the ABA problem. With this motivation, a PlusCal/TLA+ model [19] of the program was written to verify the intended solution. The PlusCal program required 103 lines to model the lock-free code. And since it produced an execution trace for the bug only with respect to the PlusCal model, it was not possible to directly employ GDB to analyze the execution trace.

Nevertheless, even with the added confidence due to the PlusCal program, the first author continued to worry about the possibility of human error in translating such snippets of DMTCP code to PlusCal. This was the motivation of the first author in investigating SimGrid, and then developing a collaboration to develop Sthread.

As will be seen in section 5.5, the analysis of the ABA problem encountered in DMTCP is considerably simpler when using Sthread, as compared to PlusCal.

## 5.5 ABA Bug

Next, the code for the lock-free algorithm for memory allocation, as described in section 5.4, is analyzed.

■ **Listing 5** Code snippet for jalloc.cpp program of DMTCP for lock-free memory allocation

```cpp
#include <pthread.h>
#include <stdio.h>
#include <unistd.h>
#include <signal.h>
#include <sthread.h>

static bool _initialized = false;

#include <sys/mman.h>
#include <stdlib.h>

inline void* _alloc_raw(size_t n)
{
  void* p = mmap(NULL, n, PROT_READ |
    ↪ PROT_WRITE, MAP_PRIVATE |
    ↪ MAP_ANONYMOUS, -1, 0);
  if(p==MAP_FAILED) { perror("_alloc_raw: "); }
  return p;
}

inline void _dealloc_raw(void* ptr, size_t n)
{
  if(ptr==0 || n==0) return;
  int rv = munmap(ptr, n);
  if(rv!=0)
    perror("_dealloc_raw: ");
}

template <size_t _N>
class JFixedAllocStack {
public:
  enum { N=_N };
```

```cpp
JFixedAllocStack() {
  if (_blockSize == 0) {
    _blockSize = 10*1024;
    _root = NULL;
  }
}

void initialize(int blockSize) {
  _blockSize = blockSize;
}

size_t chunkSize() { return N; }

//allocate a chunk of size N
void* allocate() {
  FreeItem* item = NULL;
  do {
    if (_root == NULL) {
      expand();
    }

    // NOTE: _root could still be NULL (if
    //     other threads consumed all
    // blocks that were made available
    //     from expand(). In such case,
    //     we
    // loop once again.

    /* Atomically does the following
    //     operation:
    * item = _root;
    * _root = item->next;
    */
    sched_yield(); // _root is a shared
    //     variable
    item = _root;
    sched_yield(); // root, item and item->
    //     next are shared
  } while (!_root || !
    //     __sync_bool_compare_and_swap
    //     (&_root, item, item->next));

  item->next = NULL;
  return item;
}

//deallocate a chunk of size N
void deallocate(void* ptr) {
  if (ptr == NULL) return;
  FreeItem* item = static_cast<FreeItem*>(
    //     ptr);
  do {
    /* Atomically does the following
    //     operation:
    * item->next = _root;
    * _root = item;
    */
    sched_yield(); // _root is a shared
    //     variable
    item->next = _root;
    sched_yield(); // root, item and item->
    //     next are shared
  } while (!
    //     __sync_bool_compare_and_swap
    //     (&_root, item->next, item));
}

protected:
  //allocate more raw memory when stack is
  //     empty
  void expand() {
    FreeItem* bufs = static_cast<FreeItem*>(
      //     _alloc_raw(_blockSize));
    int count= _blockSize / sizeof(FreeItem);
    for(int i=0; i<count-1; ++i){
      bufs[i].next=bufs+i+1;
    }

    do {
      /* Atomically does the following
      //     operation:
      * bufs[count-1].next = _root;
      * _root = bufs;
      */
      bufs[count-1].next = _root;
    } while (!
      //     __sync_bool_compare_and_swap
      //     (&_root, bufs[count-1].next, bufs)
      //     );
  }

protected:
  struct FreeItem {
    union {
      FreeItem* next;
      char buf[N];
    };
  };
private:
  FreeItem* volatile _root;
  size_t _blockSize;
  char padding[128];
};
```

```
// The original code had 4 levels of blocks of
       ↪ different sizes.
JFixedAllocStack<64> lvl1;

void initialize(void)
{
  lvl1.initialize(1024*16);
  _initialized = true;
}
void* allocate(size_t n)
{
  if (!_initialized) {
    initialize();
  }
  void *retVal;
  if(n <= lvl1.chunkSize()) retVal = lvl1.allocate
       ↪ (); else
  retVal = _alloc_raw(n);
  return retVal;
}
void deallocate(void* ptr, size_t n)
{
  if (!_initialized) {
    char msg[] = "***DMTCP INTERNAL ERROR:
         ↪ Free called before init\n";
    abort();
  }
  if(n <= lvl1.N) lvl1.deallocate(ptr); else
  _dealloc_raw(ptr, n);
}
```

```
// This function added for testing
void *allocate_tester() {
  // Each thread has a private item array
  void *item[20];
  int idx = 0;
  for (int i = 0; i < 5; i++) {
    if ((random() & 1) == 1) {
      item[idx++] = lvl1.allocate();
    } else if (idx > 0) {
      lvl1.deallocate(item[--idx]);
    } // Else do nothing
  }
  return NULL;
}

int main(int argc, char* argv[])
{
  // Initialize 1024 blocks of 16 bytes
  initialize();
  // Create second thread, with competing
       ↪ calls by allocate_tester()
  srandom(42);
  pthread_t thread2;
  pthread_create(&thread2, NULL,
       ↪ allocate_tester, NULL);
  allocate_tester();
  return 0;
}
```

The code initializes a shared stack of 10,240 64-byte items. The two threads concurrent allocate and deallocate items from the shared stack. Each thread allocates and deallocates at random, but the POSIX-standard sthread() call is used to initialize the random seed, so that repeated runs will return the same pseudo-random sequence.

```
[ 0.000000] (0:maestro@) Check a safety property. Reduction is: none.
[ 0.000000] (0:maestro@) Configuration change: Set 'model-check/reduction' to 'none'
Segmentation fault.
[ 0.000000] (0:maestro@) **************************
[ 0.000000] (0:maestro@) ** CRASH IN THE PROGRAM **
[ 0.000000] (0:maestro@) **************************
[ 0.000000] (0:maestro@) From signal: Segmentation fault
[ 0.000000] (0:maestro@) A core dump was generated by the system.
[ 0.000000] (0:maestro@) Counter-example execution trace:
[ 0.000000] (0:maestro@) Path = 2;2;2;2;1;2;2;1
[ 0.000000] (0:maestro@) Expanded states = 14155
[ 0.000000] (0:maestro@) Visited states = 90784
[ 0.000000] (0:maestro@) Executed transitions = 89703
[ 0.000000] (0:maestro@) Stack trace:
  0: JFixedAllocStack<64ul>::allocate() (…)
  1: allocate_tester() (…)
  2: primary_thread(int, char**) (…)
```

```
 3: int std::__invoke_impl<int, int (*&)(int, char**), … (…)
 4: std::__invoke_result<int (*&)(int, char**), int&, … (…)
 5: int std::_Bind<int (*(int, char**))(int, char**)>:: …
 6: int std::_Bind<int (*(int, char**))(int, char**)>:: …
 7: std::_Function_handler<void (), std::_Bind<int …
 8: std::function<void ()>::operator()() const (…)
 9: simgrid::kernel::context::Context::operator()() (…)
10: simgrid::kernel::context::RawContext::wrapper(…) (…)
11: ? (RIP=0x0 RSP=0x55555e123000)
```

Here, Sthread is faithful to the example in the DMTCP project that brought this to our attention. In the DMTCP project, the target application was observed to crash, and Sthread has discovered a thread schedule leading to a crash. Sthread discovered a program crash using its default max-depth of 1000.

The reported thread schedule 2;2;2;2;1;2;2;1 corresponds to an allocate, another allocate and a deallocate by thread 2, while thread 1 is trying to allocate. The pattern corresponds to the ABA problem earlier described in section 5.4. The choice of calls to allocate or deallocate was verified by printing application source line numbers where there are calls to sched_yield().

Finally, during deallocate, a thread has a choice of which of the buffers that it has allocated that it wishes to deallocate back to the stack. The testing functions called by main() (and not shown here) includes a standard model-checking "choose()" function, which is used to try each choice of buffers. If a choice does not lead to a bug, then this is followed by a rollback to the next choice of stack buffer to be deallocated. The application prints the choice, along with the source line numbers described above, in order to reproduce the bug in combination with the thread schedule.

Sthread terminated its check at the end of the thread schedule, 2;2;2;2;1;2;2;1 because it reached a failed assertion in the application source code at the end of allocate(). The assertion was not in the original source code, but we wished to limit the max-depth to a small integer for the sake of efficiency. So, we added assertions to the end of allocate() and deallocate() in order to check for consistency of the stack data structure. Without this, one would have to explore to a much larger max-depth in order to (a) see an inconsistent stack data structure being create; and then (b) see the application eventually crash due to the inconsistency. Further, the inconsistent data structure is a bug even if the process never crashes.

The assertion that was added follows the linked lists from _root through dereferencing (via item->next) to eventually reach the end of stack. In the code, when an item is removed from the stack, the original code is modified to set the next item to a fixed, but randomly chosen 64-bit double word. This allows for detection of inconsistencies in the case that the removed item remains accessible from the data structure. We also terminate the dereference checking after at most 64 iterations. Any dereferencing through 64 or more iterations is considered an assertion failure, since we limited the testing function to at most five calls to allocate and deallocate..

Finally, note that in case of a bug, the assertion will either produce an assertion failure, or it will produce a program crash due to a bad pointer dereference (e.g., dereferencing NULL). In our experiment, the program crashed.

## 6 Related Work

The Sthread implementation is based on SimGrid [21, 7]. SimGrid is a long-standing project that appeared in 2003 [20]. It is a simulation package that offers explicit-state model checking for formal verification. The process memory is considered as the state of the process. The SimGrid developers call this *stateless model checking* since the model state is not derived from the target program variables themselves. The state of all of process memory serves as the model-checking state. SimGrid includes sophisticated heuristics to determine what parts of process memory represent garbage and should not be included in a test for state equality. SimGrid has been used for verification of large-scale distributed systems [6], network simulations [32], and MPI [11]. The last example includes the SMPI interface, which was the inspiration for the Sthread interface of the current work.

The use of process memory as model state is the key to using the program as the formal model, without the need for the programmer to design a separate intermediate formal model for verification. The idea of using process memory as state in a model has also been used in other approaches to formal verification. An early approach to combining explicit-state model checking with process memory as state is found in CMC [24], where it was used for event-driven networking protocols in a model of the AODV distributed protocol. Rather than use all of process memory as a model state, some approaches employ a compiler to parse the target code and instrument the addresses of all variables in the code (e.g., see CBMC [9, 18], described below). In this case, the bits of memory of those (global and local) variables become the portion of memory representing the model state. In related work, [23] describes transformations of process memory (i.e., reordering of memory regions) that can be used to decide when the memory state for two processes should be considered semantically equivalent.

There is one other well-known approach that often employs the idea of modeling the memory of a process. This is the use of *Context-Bounded Analysis* (CBA) in formal verification. However, the majority of CBA approaches use a compiler such as LLVM to instrument the addresses of variables of the target program. The direct use of program variables as state facilitates such back ends as SAT-based symbolic algorithms (testing for arbitrary program input) and *Bounded Decision Diagrams* (BDDs) [1].

The general goal of CBA is different from that of Sthread. CBA methods typically have the more ambitious goal of capturing buffer overruns, writes to arbitrary addresses in memory, bugs in memory allocators, writes after freeing a memory region, and other bugs often associated either with low-level code of with C/C++ and other "unsafe" languages.

Hence, while the goal of CBA methods is to formally model *preemptive context switching*. This has the advantage of being more general than Sthread. But it also has the disadvantage of incurring a still greater exponential explosion of states than Sthread, since thread context switches are assumed to be possible at any point at all in the code. In order to capture this more ambitious goal, the CBA methods employ a source-to-source transformation of a multithreaded program into a sequential program, in which a non-deterministic program variable, current_thread, indicates which thread executes the next statement. After each statement, a "goto" is executed to go to pc[current_thread] and that pc is then incremented. Thus, any thread can preempt after any statement. An example of this approach is Lazy-CSeq [16, 17].

In order to limit the exponential explosion, the number of context switches (changes to current_thread) is bounded. (Hence, this motivates the name Context-Bounded Analysis for CBA.) In some cases, the upper bound on context switches in CBA is even set to two. A successful example of limiting the bound to two is found in KISS [29]. The implicit assumption behind a small bound on the number of context switches is that "errors in multithreaded software typically have shortest counterexample traces that require only a small number of context switches" [34]. This is especially the case for finding implementation bugs due to memory overruns, write-after-free, etc.

At the back end, CBA typically uses either a SAT solver or a Boolean Decision Diagram (BDD) to implement formal verification. In the case of symbolic-algorithm approaches, BDDs are especially popular. CBMC is an example of a popular bounded model checker for C [9, 18]. It derives its state using a compiler to identify all variables, and can unwind loops to a certain bound, and it uses SAT or SMT for the back end. JCBMC [27] uses a symbolic algorithm on Java codes in which the model state is the state of all variables.

There are two notable examples of prior work which use an explicit-state model checker as a back end, and also directly operate on C/C++ programs, and for which process memory regions can be used as part of a formal model state. One is pancam, which uses the SPIN [15] back-end for traditional explicit-state model checking. The second one is DIVINE, which (unlike Sthread) does not support programmer code for specific liveness properties. DIVINE is centered around general LTL formulas and the corresponding Büchi automata, along with explicit-state model checking. DIVINE prefers general liveness properties rather than the application-specific liveness properties of Sthread. In both cases, virtual machines interpret an intermediate representation, byte code generated by the LLVM compiler. In both cases, the use of an interpreter implies slower execution of the target program during model checking. Hence, it is not in vivo. Further, while Sthread uses all of process memory as model state, and then subtracts memory that may be garbage (e.g., uninitialized malloc memory), these approaches begin with only the memory of the program variables, and then (in the case of DIVINE), the interpreter tracks additional memory that is allocated by the target program. See [13] for a classification of model checking that includes some of the ways that model state may be represented.

In the case of CBA, pancam [34] has employed CBA with the well-known SPIN explicit-state model checker [34] at the back end. LLVM byte code is interpreted in a virtual machine and the LLVM byte code allows the system to track program variables as its state.

Another notable approach is DIVINE [3, 4, 30], a large, long-standing project now in its fourth version [3]. The citation [30] is particularly recommended for an overview of DIVINE. DIVINE takes the approach of no annotations to the target C/C++ code. Thus, DIVINE includes a large number of *general* LTL-based properties that must be checked for the entire program. (In contrast, Sthread allows the programmer to add program-specific C/C++ variables to express program-specific liveness properties to be checked. See section 5.3.)

While DIVINE directly checks input C/C++ programs, it follows the approach of pancam in compiling to LLVM byte code (IR, or intermediate representation), where it is interpreted by the DIVM virtual machine (in version 4, whereas formerly the interpreter was integrated into the formal verification layer). While the capabilities of DIVINE are impressive, it suffers from two failings with respect to the goals of this paper: (i) the use of an interpreter for intermediate byte code means that it is slower than Sthread's execution of native machine code, and the DIVINE interpreter is limited in not being able to handle most kernel calls or any external calls to a GPU, FPGA or other CPU accelerator [30, page 2]; and (ii) the pre-compilation phase to compile *general* LTL formulas results in larger Büchi automata, which are both CPU-intensive and consume a lot of memory. Concerning issue (i), note that: "DIVINE can often be directly applied to verification of real-world code, provided it does not use inputs or platform-specific features, such as calls into the kernel of the operating system" [30, page 2]. This also implies that extensions of DIVINE to other communication libraries would imply a major undertaking, while Sthread has already demonstrated its in-vivo capability in SimGrid/SMPI for MPI [11]. Concerning issue (ii), DIVINE reports that its multi-core OWCTY plus on-the-fly cycle detection heuristic relies on topological sort for cycle detection and is "quadratic in the worst case for general LTL properties, although for a significant subset of formulae ([for] weak Büchi automata) the algorithm runs in linear time in the size of the product automaton" [12, see DiVinE-Multi-Core Algorithm]. In comparison, Sthread directly tests for properties such as deadlock, and can use simpler LTL formulas to test liveness since it allows the programmer to annotate the target program with application-specific livelock properties of concern only to that program. In fairness to DIVINE, issue (i) (interpreted execution) is partially mediated by DIVINE's re-implementation of the libc and pthread libraries for compatibility with DIVINE's interpret. Issue (ii) (precomputation of general LTL properties with Büchi automata) is partially mediated by DIVINE's use of multi-core and clusters for compilation of the Büchi automata [5].

An early example related to the CBA approach (although not known under that name at the time) is Chess [25]. Chess modeled multithreaded system calls, but it was not formally a model checker since it did not have states. It simply explored many thread schedules from the beginning and had neither process memory-based states or other states. $S^2E$ [8] is related to Chess, in that it is not a model checker since it does not have states. It employs a symbolic algorithm with abstractions to form a tree of search schedules, and it refines that tree to explore additional execution paths. It operates directly on code by employing dynamic binary translation inside a virtual machine.

In contrast to the above work, the current work is intended to be employed in concurrent algorithms in which a larger number of context switches are required to exhibit a bug (as opposed to the case for CBA).

## 7 Conclusion and Future Work

It has been shown that Sthread correctly handles a broad variety of concurrent algorithms for multithreaded programs. Examples were shown for deadlock, program crash, assertion failure, and livelock. Sthread allows the programmer to add small in-vivo annotations in C/C++ to the target program. This provides a reasonable balance between allowing the programmer to provide hints to Sthread for better efficiency, and maintaining a light burden for programmers with no experience in formal verification.

The examples covered here also strongly overlap with the exercises in a beginning course covering multithreaded programming. This is meant to suggest the pedagogical advantages of using Sthread to allow beginning students to gain immediate feedback through execution traces of failure modes, as they seek to debug their homework.

*Dynamic partial order reduction* (DPOR) is a classical technique used to make explicit-state model checking more efficient. SimGrid already employs DPOR in other domains, and SimGrid will be extended to support DPOR for the POSIX threads domain.

While the work of this paper is based on the C/C++ language, Sthread is both language-independent and compiler-independent. It operates directly on machine code, rather than a compiler-generated intermediate byte code. This will be demonstrated in future work, by extending the Sthread implementation to support verification of multithreaded APIs for other languages.

In order to better support pedagogy, a future work will develop a series of graduated exercises in implementing POSIX threads programs and to implement concurrent algorithms with multithreading. A framework will be designed to automatically employ Sthread to check the student's work and provide feedback, rather than requiring the student to directly employ Sthread, as is currently the case.

**Acknowledgements** The work of the first author was partially supported by NSF Grant OAC-1740218. The work of the second author was partially supported by the Inria Project Lab (IPL) Hac Specis. The two together are partially supported by a grant from Inria as an Inria Associate Team (Équipe Associée d'Inria) with the title "FogRein: Steering Efficiency for Distributed Applications". The authors are grateful for the detailed comments of the referees and for the careful reading and comments by Gregory Price.

**Sthread: In-Vivo Model Checking of Multithreaded Programs**

**About the authors**

**Gene Cooperman** Contact him at gene@ccs.neu.edu.

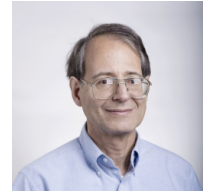

**Martin Quinson** Contact him at martin.quinson@ens-rennes.fr.

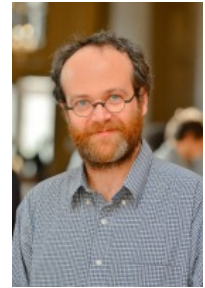